# P-type Ohmic contact to monolayer WSe$_2$ field-effect transistors using high electron affinity amorphous MoO$_3$

*Yi-Hsun Chen[1,2]†, Kaijian Xing[1]†, Song Liu[3], Luke Holtzman[4], Daniel L. Creedon[5], Jeffrey C. McCallum[5], Kenji Watanabe[6], Takashi Taniguchi[7], Katayun Barmak[4], James Hone[3], Alexander R. Hamilton[8], Shao-Yu Chen[2,9]\*, Michael S. Fuhrer[1,2,10]\**

[1] School of Physics and Astronomy, Monash University, Clayton, Victoria 3800, Australia

[2] Australian Research Council Centre of Excellence in Future Low-Energy Electronics Technologies (FLEET), Monash University, Clayton, Victoria 3800, Australia

[3] Department of Mechanical Engineering, Columbia University, New York, New York 10027, United States

[4] Department of Applied Physics and Applied Mathematics, Columbia University, New York, New York 10027, United States

[5] School of Physics, the University of Melbourne, Melbourne, VIC 3010, Australia

[6] Research Center for Functional Materials, National Institute for Materials Science, 1-1 Namiki, Tsukuba 305-0044, Japan

[7] International Center for Materials Nanoarchitectonics, National Institute for Materials Science, 1-1 Namiki, Tsukuba 305-0044, Japan

[8] School of Physics, University of New South Wales, 2052 Sydney, Australia

[9] Center of Condensed Matter Sciences and Center of Atomic Initiative for New Material, National Taiwan University, Taipei 106, Taiwan

[10] Monash Centre for Atomically Thin Materials, Monash University, Clayton, 3800, VIC, Australia

**ABSTRACT**

Monolayer tungsten diselenide (1L-WSe$_2$) has been widely used for studying emergent physics due to the unique properties of its valence bands. However, electrical transport studies have been impeded by the lack of a reliable method to realize Ohmic hole-conducting contacts to 1L-WSe$_2$. Here, we report low-temperature p-type Ohmic contact between 1L-WSe$_2$ and molybdenum trioxide (MoO$_3$) where 1L-WSe$_2$ underneath MoO$_3$ is heavily hole doped through surface transfer doping. Electrical transport measurements reveal linear current-voltage relations at 10 K and nearly temperature-independent output curves up to room temperature, which support our finding. Furthermore, the contact resistivity of MoO$_3$-contacted 1L-WSe$_2$ FET is 25.9–66.2 kΩ·μm at a low carrier density of $1.5 \times 10^{12}$ cm$^{-2}$, which is the lowest reported for 1L-WSe$_2$ FETs. Realizing robust p-type Ohmic contact to a 2D transition metal dichalcogenide semiconductor will advance the development of microelectronics based on correlated phenomena in the valence bands.

Monolayer transition metal dichalcogenides (TMDs) have been extensively studied for their profound electronic and optoelectronic properties.[1–5] Among the TMDs, monolayer WSe$_2$ (1L-WSe$_2$) is promising for investigating emergent physics arising in the valence band, owing to its low ionization potential[6–8] and strong spin-orbit coupling resulting in fully spin-polarized valleys with large spin-splitting of 0.4 V.[9,10] 1L-WSe$_2$ has been used to study correlated physics[11–15], resonant tunneling[16], and Bose-Einstein condensation[17,18] in van der Waals heterostructures. However, studying these exotic phenomena through electronic transport often necessitates other techniques, such as capacitive[19,20] or optical sensing[12,15], due to the difficulty in forming p-type Ohmic contacts to 1L-WSe$_2$. This issue becomes exponentially challenging at low temperatures.

A common strategy to fabricate Ohmic contact in conventional semiconductor devices is to dope the material underneath the contact heavily, for example, by ion implementation and subsequent high-temperature annealing[21]. However, such methods are not appropriate for atomically-thin van der Waal materials where the damage generated during implantation is hard to repair and can cause severe Fermi-level pinning[22]. Instead, surface transfer doping offers a favorable alternative, as it creates minimal defects in the material. Recently, surface transfer doping with transition metal oxides, such as $MoO_3$, $WO_3$, and $V_2O_5$, has been used to achieve high levels of p-doping in hydrogenated diamond[23,24] and TMD[25,26] based field-effect transistors (FETs).

In this work, we use amorphous $MoO_3$ as a surface transfer dopant to realize p-type Ohmic contacts to 1L-WSe$_2$ FETs. Electronic transport measurements on $MoO_3$-contacted 1L-WSe$_2$ FETs reveal Ohmic behavior, supported by linear and nearly temperature-independent output curves from room temperature to 10 K. We find that the contact resistance ($R_c$) is about 1.8–4.6 kΩ (contact resistivity 25.9–66.2 kΩ·μm) at a carrier density of $1.5 \times 10^{12}$ cm$^{-2}$, the lowest among those reported for 1L-WSe$_2$ FETs in hole transport regime. Examination of the activation barriers for bulk and contacts shows that activated behavior is dominated by the bulk even for carrier ensities into the subthreshold regime. The realization of p-type Ohmic contacts enables high-quality 1L-WSe$_2$ FETs and further paves the way to developing quantum electronic devices based on correlated quantum phenomena in WSe$_2$ valance bands.

Figures 1a and 1b show schematically the relative positions of the electronic bands of MoO$_3$ and 1L-WSe$_2$ and the resulting junction at the contact regime in equilibrium. The electron affinity of MoO$_3$ (6.7 eV)[27] is higher than the ionization potential of 1L-WSe$_2$ (4.8–5.2 eV)[6–8], resulting in p-doping of the WSe$_2$. Here we consider MoO$_3$ as a highly-doped n-type semiconductor, while some previous works have treated it as a metal with a high work function[28,29]. The spontaneous charge transfer between MoO$_3$ and 1L-WSe$_2$ increases the electron (hole) density in MoO$_3$ (1L-WSe$_2$), leading to a non-rectifying barrier with no depletion region at the interface between the doped contact region and the pristine channel (Figure 1b). Moreover, MoO$_3$ has a lower deposition temperature of about 475°C, compared to traditional high work function metals (Pd, Pt, and Au), which may moderate the impacts of evaporated atoms on the 1L-WSe$_2$ and maintain a high-quality interface by minimizing disorder-induced gap states.[30,31]

We first confirm surface transfer doping by performing low-temperature photoluminescence (LT-PL). Figures 1c and 1d show the LT-PL spectra of 1L-WSe$_2$ at 80 K before and after depositing MoO$_3$, respectively. In Figures 1c and 1d, the PL spectrum is composed of two distinct peaks: the high-energy one is from the neutral excitons, and the lower-energy one is from trions (a neutral exciton bound to a free charged carrier). For pristine 1L-WSe$_2$ (Figure 1c) the energy separation of these two peaks is about 34 meV, corresponding to the binding energy of negative trions (X$^-$, exciton bound to a free electron).[32] The presence of X$^-$ indicates the n-type doping of as-prepared 1L-WSe$_2$. In contrast, after depositing MoO$_3$ (Figure 1d), the binding energy of trions changes to 18 meV, consistent with positive trions (X$^+$, exciton bound to a free hole).[32,33]. The significant alteration of trion species from negative to positive trions demonstrates the significant p-type surface transfer doping of 1L-WSe$_2$ by MoO$_3$.

Figure 2a shows the optical micrograph of a MoO$_3$ contacted 1L-WSe$_2$ FET with hBN encapsulation. The detailed fabrication processes can be found in Supporting Information S1. Briefly, 1L-WSe$_2$/hBN heterostructures were firstly fabricated using WSe$_2$ crystals grown by the self-flux method[34]. After that, we deposited 20 nm MoO$_3$ and 60 nm Pd in sequence through typical photolithography, then transferred a top hBN layer to improve the sample stability. We note that hBN encapsulation effectively reduces extrinsic Coulomb scatterers from both ambient and the SiO$_2$/Si substrate[35]. In addition, depositing Pd can enhance electrical conduction and further protect MoO$_3$ from degradation; the work function of MoO$_3$ decreases when exposed to the ambient due to the reduction of oxygen vacancies[36]. The cross-sectional schematic of MoO$_3$ contacted 1L-WSe$_2$ FETs is shown in Figure 2b. Control devices were fabricated similarly with electrodes of 60 nm Pd only.

Figure 3 shows the output characteristics of a MoO$_3$ contacted 1L-WSe$_2$ FET and the control device at various gate voltages ($V_{gs}$) and temperatures. The measurements are conducted by applying a source-drain voltage ($V_{ds}$) across two outer electrodes and simultaneously measuring the drain current ($I_{ds}$) and the four-probe voltage ($V_{4p}$). Figure 3a shows low-bias ($|V_{ds}| \leq 100$ mV) output curves of the MoO$_3$ contacted 1L-WSe$_2$ FET at $T = 10$ K for different $V_{gs}$. It exhibits a typical behavior of hole conduction, where the threshold voltage ($V_{th}$) is estimated as −68 V (see Figure S3a). The output curves are linear and symmetric for a wide range of $V_{gs}$, suggesting robust Ohmic behavior of MoO$_3$-contacted 1L-WSe$_2$ even at such low temperature. Notably, the linear current-voltage relationship can be observed even in the subthreshold region ($V_{gs} >$ −68 V). The linear output curves indicate any energy barrier at the metal-semiconductor interface smaller than a few times the thermal energy $k_BT = 0.86$ meV, where $k_B$ is the Boltzmann constant. As a comparison in Figure 3b, the control device exhibits nonlinear output curves at $T$ = 10 K with a much lower output current at the same $V_{gs}$, suggesting that a sizable energy barrier limits the conduction. Since the devices were fabricated identically except for the contacts, we attribute the outstanding output characteristics in the $MoO_3$ contacted 1L-$WSe_2$ FET to the Ohmic contact formed at the $MoO_3$/1L-$WSe_2$ interface.

Figures 3c and 3d show temperature-dependent output curves at $V_{gs}$ = −80 V for both devices. The $MoO_3$ contacted 1L-$WSe_2$ FET exhibit linear output curves with little temperature dependence over a wide temperature range ($T$ = 10–250 K), consistent with the observation of a negligible activation barrier. This is significant in contrast to the control device, whose output curves become more resistive and highly nonlinear below $T$ = 100 K, indicating that considerable thermal energy is required to overcome the Schottky barrier. We note that the conductivity of the $MoO_3$-contacted 1L-$WSe_2$ FET slightly decreases below $T$ = 150 K. This might be due to the temperature dependence of $V_{th}$ (Supporting Information S4): The $V_{th}$ shifts negatively when temperature decreases, resulting in a reduction of induced free carrier density in $WSe_2$.

We quantitatively model the current flow in our $MoO_3$ contacted 1L-$WSe_2$ FETs in order to extract $R_c$. Figure 4a shows the two-probe conductance $G_{2p} = I_{ds}/V_{ds}$ and four-probe conductance $G_{4p} = I_{ds}/V_{4p}$ as a function of $V_{gs}$ at $T$ = 77 K. Both curves present a typical turn-on behavior with $V_{th}$ = −58 V (see Figure S3b). The $G_{4p}$ is about 3 times higher than $G_{2p}$, consistent with the ratio of the channel length: 19 μm for $V_{ds}$ probes and 7 μm for $V_{4p}$ probes. Furthermore, we found that the two-probe ($μ_{2p}$) and four-probe mobilities ($μ_{4p}$) extracted from two- and four-terminal measurements are almost identical: $μ_{2p}$ is 56.1 cm$^2$/V·s and $μ_{4p}$ is 61.2 cm$^2$/V·s at $|V_{gs} − V_{th}|$ = 17 V. Both results imply that the $MoO_3$-contacted devices exhibit small $R_c$, consistent with the previous discussion in Figure 3. Note that we are only able to present $G_{4p}$ at $V_{gs} \leq$ −33 V because of instrumental limitation when the channel becomes highly resistive in our setup.

Typically, $R_c$ can be extracted from four-terminal measurements by calculating

$$2R_c = R_{2p} - \alpha R_{4p} \qquad (1),$$

where $R_{2p}$ and $R_{4p}$ are the two- and four-probe resistance, respectively; α is geometry factor. For an ideal Hall bar geometry, $\alpha = L_{2p}/L_{4p}$ can be treated as a constant and $R_c$ can be calculated straightforwardly using Eqn. (1), where $L_{4p}$ is the length between $V_{4p}$ electrodes and $L_{2p}$ is the length between $V_{ds}$ electrodes. However, it is critical to consider the device geometry in the case when the $V_{4p}$ probes are invasive. For instance, some current may flow through the electrodes, making the determination of α uncertain. Moreover, α may depend on contact resistance as it affects the relative distribution of current flowing through the sample and electrodes.

To tackle the above issues, we determine $R_c$ by performing finite-element modelling by COMSOL simulation (see Supporting Information S6). We considered a two-dimensional model where electrodes ($MoO_3$/Pd in our case) are embedded in the channel. As illustrated in Figures S6a and S6b, the device geometry is determined by the nominal dimensions of the photolithography pattern. The TMD-metal contacts can be modelled by a width of $d$ within the nominal electrode boundary. These marginal regions account for both "transfer length" under electrodes and any dimensional differences from nominal patterns caused by under/overexposure. In addition, we assigned the conductivity of channel material ($\sigma_{channel}$), contact regions ($\sigma_c$), and metal ($\sigma_m$) independently. $\sigma_{channel}$ can be approximated from the two-terminal conductance $\sigma_{channel} = G_{4p} \times \frac{L_{2p}}{W}$, where $W$ is the width of the channel, since the contact resistance is small in the $MoO_3$ contacted 1L-$WSe_2$ FET. $\sigma_m$ is approximated by the conductivity of Pd: $\sigma_m = 4.8 \times 10^5$ S. With this model, $R_c$ can be calculated as $R_c = \frac{1}{\sigma_c} \frac{W}{d}$ given $\sigma_c$ and $d$. We cannot determine $\sigma_c$ and $d$ unambiguously from a single measurement of the ratio of $V_{4p}/V_{ds}$ at each $V_{gs}$ and $T$. However, $d$ should be smaller than one-half the electrode width and cannot be significantly negative (corresponding to actual electrodes wider than nominal dimensions) as this would be apparent in optical microscopy. Fortuitously, we find in our calculations that $d$ must be at least a small positive value (e.g. 183 nm at $V_{gs} = -80$ V and $T = 77$ K), and that the $\sigma_c/\sigma_{channel}$ for which $V_{4p}/V_{ds}$ matches the experiment does not depend strongly on $d$. Therefore, we are able to strongly bound the values of $\sigma_c$ (and hence $R_c$) that are consistent with the measurement.

Figure 4b shows calculated $R_c$ as a function of temperature at $V_{gs}$ ranging from $-60$ to $-80$ V. $R_c$ is strongly dependent on $V_{gs}$ but only weakly dependent on temperature, as expected for a negligible activation barrier. At $T = 77$ K, we found $R_c$ approximately 1.8–4.6 k$\Omega$ at $V_{gs} = -80$ V. According to the conventional parallel plate capacitance approximation, $n = C_{ox}|V_{gs} - V_{th}|/q$, the corresponding carrier density ($n$) at $V_{gs} = -80$ V is $1.5 \times 10^{12}$ cm$^{-2}$ ($|V_{gs} - V_{th}| = 22$ V). It is worth noting that this is the first work to demonstrate p-type Ohmic contact to 1L-WSe$_2$ at a low carrier density ($n = 1.5 \times 10^{12}$ cm$^{-2}$). The $R_c = 1.8$–4.6 k$\Omega$ (or contact resistivity 25.9–66.2 k$\Omega\cdot\mu$m) is favorable compared to previously reported 1L-WSe$_2$ FETs,[37] and comparable to Co/hBN tunneling contact in monolayer MoS$_2$[38]. The $R_c$ dependence on temperature is also sensitive to different $V_{gs}$. At a small negative $V_{gs}$, $R_c$ is weakly sensitive to temperature. However, $R_c$ is almost independent of temperature at a high $V_{gs}$, such as $V_{gs} = -80$ V. This can be attributed to a higher conductivity of 1L-WSe$_2$ underneath MoO$_3$ while applying a high $V_{gs}$.

In Figure 4c, we compare $R_c$ by two methods mentioned above: COMSOL simulation and Eqn. (1). We set $\alpha = 2.7$ determined by the ratio of the physical distances between the $V_{ds}$ and $V_{4p}$ electrodes to calculate $R_c$ based on Eqn. (1). We found the result acquired by the naïve application of Eqn. (1) overestimates $R_c$ about 2 times at $V_{gs} = -80$ V and underestimates $R_c$ at $V_{gs} \geq -77$ V compared to the simulation. The failure of Eqn. (1) results from the assumption that the device has an ideal Hall bar geometry with a constant $\alpha$, leading to an incorrect estimation of $R_c$ when it is applied to a device with invasive $V_{4p}$ electrodes. Our results emphasize the significance of considering the geometry factor to estimate $R_c$ in a device with non-ideal geometry.

We provide further evidence of an Ohmic contact between $MoO_3$-doped $WSe_2$ and the $WSe_2$ channel by measuring the activation barriers for the contacts and the channel as a function of $V_{gs}$ separately. We estimate the activation barrier of the contacts $\Phi_{2p}$ using the standard thermionic emission model[39]:

$$\sigma_{ds} = A^* T^{3/2} \exp\left[-\frac{\Phi_{2p}}{k_B T}\right]\left[1 - \exp\left(\frac{qV_{ds}}{k_B T}\right)\right],$$

where $A^*$ is Richardson's constant, $q$ is the elementary charge, $k_B$ is the Boltzmann constant, and $T$ is the temperature. $\Phi_{2p}$ is the energy required to overcome the contact barrier *via* the thermionic emission process, which can be extracted by fitting the slope in the Arrhenius plot at each $V_{gs}$. We estimate the activation barrier of the $WSe_2$ channel by assuming that the four-probe conductivity $\sigma_{4p}$ is thermally activated, i.e. $\sigma_{4p} \propto e^{-\Phi_{4p}/k_B T}$, where $\Phi_{4p}$ is the activation energy (energy difference between Fermi energy and valence band edge or mobility edge) in the $WSe_2$ channel.

Figures 5a and 5b show Arrhenius plots of $\sigma_{4p}$ and $\sigma_{ds}/T^{3/2}$ for the $MoO_3$ contacted 1L-$WSe_2$ FET, respectively. The $\Phi_{4p}$ is 22 meV at $V_{gs} = -34$ V and then decreases to 3 meV at $V_{gs} = -80$ V, as shown in Figure S5a. The point at which $\Phi_{4p}$ becomes smaller than thermal energy $k_B T$ (8.6 meV for $T = 100$ K) marks a transition from thermally activated (sub-threshold) behavior at $V_{gs} \geq -50$ V to relatively temperature-independent conduction at $V_{gs} \leq -50$ V, which agrees reasonably well with the estimated $V_{th}$ in Supporting Information S3. Figure 5c shows the activation barriers of bulk and contacts, $\Phi_{4p}$ and $\Phi_{2p}$, as functions of $V_{gs}$ extracted from Arrhenius plots of $\sigma_{4p}$ and $\sigma_{ds}/T^{3/2}$, respectively (Figures 5a,b). The observations are surprising, and at odds with the conventional picture in the case of a Schottky barrier at the metal-semiconductor interface[39], as illustrated in Figure 5d: At small negative $V_{gs}$ (orange lines, Figure 5d), the bulk barrier dominates conduction, and $\Phi_{4p} \approx \Phi_{2p}$. As the $V_{gs}$ is tuned more negative (blue and green lines, Figure 5d), the bulk barrier $\Phi_{4p}$ becomes smaller but the contact barrier $\Phi_{2p}$ saturates at the Schottky barrier height. Figure 5e shows that, in contrast, for the interface between a heavily p-doped semiconductor contact and a gate-controlled semiconductor channel, the condition $\Phi_{4p} = \Phi_{2p}$ is expected at the flat-band condition and $\Phi_{4p} \approx \Phi_{2p}$ is expected at $V_{gs}$ beyond the flat-band condition.

Examining the $V_{gs}$ dependence of $\Phi_{4p}$ and $\Phi_{2p}$ (Figure 5c), the observations are consistent with an Ohmic contact between a heavily p-doped semiconductor contact and the gate-controlled channel (Figure 5e). In particular, we find that $\Phi_{4p}$ is comparable to $\Phi_{2p}$ at $V_{gs} \leq -37$ V ($|\Phi_{2p} - \Phi_{4p}| \leq k_B T \approx 8.6–17.2$ meV at $T = 100–200$ K). Moreover, we never observe a regime in which $\Phi_{2p}$ saturates while $\Phi_{4p}$ decreases with increasingly negative $V_{gs}$, as typically observed for Schottky barrier contacts. At more negative $V_{gs}$, $\Phi_{2p}$ appears even to drop below $\Phi_{4p}$, and is undetectable for $V_{gs} \leq -46$ V, even though bulk activation is still evident (the device is in the subthreshold regime). This possibly reflects the smooth nature of the contact barrier (Figure 5e) or difficulty in measuring thermionic emission for $\Phi_{2p} \leq k_B T$. The practical result is that the temperature dependence due to thermionic emission at the contacts is negligible for all $V_{gs}$ above threshold, and even well into the subthreshold regime. We also measured a Pd-only contacted 1L-WSe$_2$ control device (Figure 5f), which shows clear thermal activation in a higher temperature range (200 K $\leq T \leq$ 300 K) with $\Phi_{2p}$ remaining large at $V_{gs}$ values at which the barrier in the MoO$_3$-contacted device is negligible.

In conclusion, we have demonstrated that surface transfer doping by MoO$_3$ can be used to create p-type Ohmic contacts to 1L-WSe$_2$ enabling low-temperature transport measurements. Furthermore, we have shown that finite element modeling is surprisingly useful for extracting the contact resistance in devices with a non-standard geometry such as invasive contacts, simplifying the fabrication of devices for contact resistance measurements. Our surface transfer doping strategy is simple to implement and should be broadly applicable to other van der Waals layered semiconductors, as well as heterostructures and twisted-layer structures, enabling electronic transport studies of novel correlated phenomena in the valence bands of 2D semiconductors.

## ASSOCIATED CONTENT

**Supporting Information.**

The Supporting Information is available free of charge.

S1. Device fabrication of 1L-WSe$_2$ field-effect transistors for electrical transport measurements and MoO$_3$/1L-WSe$_2$ heterostructure for photoluminescence spectroscopy

S2. Scanning tunneling spectroscopic measurement of WSe$_2$ crystals

S3. Estimation of threshold voltage for 1L-WSe$_2$ FETs

S4. Drain current as a function of $|V_{gs}-V_{th}|$

S5. Estimation of activation energy of 1L-WSe$_2$ FETs

S6. Calculation of contact resistance by using finite element method.

## AUTHOR INFORMATION

**Corresponding Author(s)**

**Michael S. Fuhrer** – School of Physics and Astronomy and Australian Research Council Centre of Excellence in Future Low-Energy Electronics Technologies (FLEET), Monash University, Clayton, Victoria 3800, Australia; ORCID iD: https://orcid.org/0000-0001-6183-2773; Email: michael.fuhrer@monash.edu

**Shao-Yu Chen** – Center of Condensed Matter Sciences and Center of Atomic Initiative for New Material, National Taiwan University, Taipei 106, Taiwan; Australian Research Council Centre of Excellence in Future Low-Energy Electronics Technologies (FLEET), Monash University, Clayton, Victoria 3800, Australia; ORCID iD: https://orcid.org/0000-0003-3423-9768; Email: shaoyuchen@ntu.edu.tw

**Authors**

**Yi-Hsun Chen** – School of Physics and Astronomy and Australian Research Council Centre of Excellence in Future Low-Energy Electronics Technologies (FLEET), Monash University, Clayton, Victoria 3800, Australia; ORCID iD: https://orcid.org/0000-0001-6313-5189

**Kaijian Xing** – School of Physics and Astronomy, Monash University, Clayton, Victoria 3800, Australia; ORCID iD: https://orcid.org/0000-0001-5254-4710

**Song Liu** – Department of Mechanical Engineering, Columbia University, New York, New York 10027, United States

**Luke Holtzman** – Department of Applied Physics and Applied Mathematics, Columbia University, New York, New York 10027, United States

**Daniel L. Creedon** – School of Physics, the University of Melbourne, Melbourne, VIC 3010, Australia

**Jeffrey C. McCallum** – School of Physics, the University of Melbourne, Melbourne, VIC 3010, Australia

**Kenji Watanabe** – Research Center for Functional Materials, National Institute for Materials Science, 1-1 Namiki, Tsukuba 305-0044, Japan

**Takashi Taniguchi** – International Center for Materials Nanoarchitectonics, National Institute for Materials Science, 1-1 Namiki, Tsukuba 305-0044, Japan

**Katayun Barmak** – Department of Applied Physics and Applied Mathematics, Columbia University, New York, New York 10027, United States

**James Hone** – Department of Mechanical Engineering, Columbia University, New York, New York 10027, United States

**Alexander R. Hamilton** – School of Physics, University of New South Wales, 2052 Sydney, Australia

**Author Contributions**

†These authors contributed equally. Y. H. Chen, K. Xing, S. Y. Chen, and M. S. Fuhrer conceived and designed this project. Y. H. Chen fabricated the devices with assistance from K. Xing with the $MoO_3$ and Pd deposition. Y. H. Chen performed the device characterization, data analysis, and finite element modeling. K. Xing provided the concept of utilizing high electron affinity $MoO_3$ to surface transfer dope 1L-$WSe_2$. S. Y. Chen performed the low-temperature photoluminescence measuemrnet. S. Liu, J. Hone, L. N. Holtzman, and K. Barmak provided the $WSe_2$ crystals. D. L. Creedon, J. C. McCallum provided the resouruces. The manuscript was written through contributions of all authors. All authors have given approval to the final version of the manuscript.

**Notes**

The authors declare no competing financial interest.

**ACKNOWLEDGMENT**

Y. H. Chen thanks Mr. Pei-Wei Chi for insightful discussion regarding the finite element method. Tungsten diselenide crystal growth was supported under the United States National Science Foundation Materials Research Science and Engineering Center through grants DMR-1420634 and DMR-2011738. Y. H. Chen, A. R. Hamilton, M. S. Fuhrer, and S. Y. Chen acknowledge support from the ARC Centre of Excellence in Future Low-Energy Electronics Technologies (FLEET; CE170100039). K. Watanabe and T. Taniguchi acknowledge support from the JSPS KAKENHI (Grant Numbers 19H05790, 20H00354 and 21H05233). S. Y. Chen acknowledges support from the Center of Atomic Initiative for New Materials, National Taiwan University (grant nos. 110 L9008 and 111 L9008), from the Featured Areas Research Center Program within the framework of the Higher Education Sprout Project by the Ministry of Education of Taiwan. K. Xing acknowledges support from ARC grant DP200101345. S. Y.

Chen thanks the experimental support from Dr. Li-Chyong Chen and Dr. Yu-Ling Liu in the Center of Condensed Matter Sciences at National Taiwan University, and Dr. Wei-Hua Wang in the Institute of Atomic and Molecular Sciences, Academia Sinica. This work was performed in part at the Melbourne Centre for Nanofabrication (MCN) in the Victorian Node of the Australian National Fabrication Facility (ANFF).

TABLE OF CONTENTS/ABSTRACT GRAPHICS

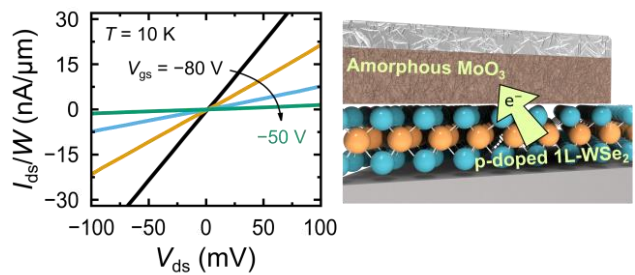

FIGURES

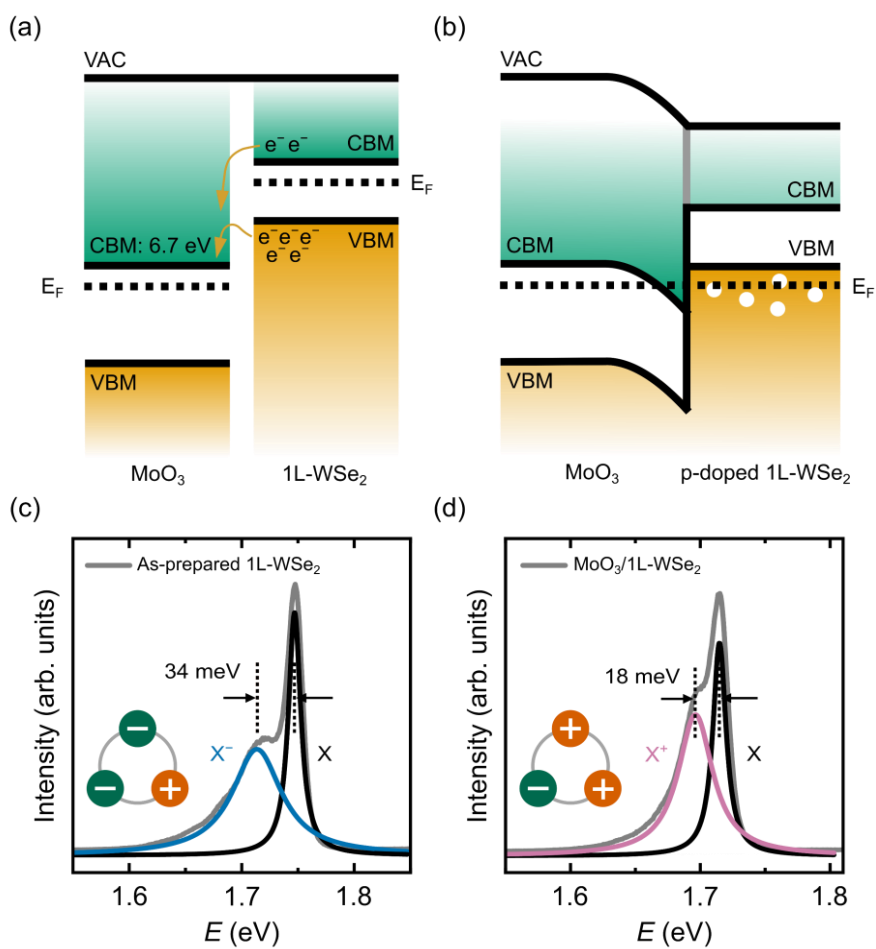

**Figure 1.** (a-b) Schematic band diagrams of isolated $MoO_3$ and 1L-$WSe_2$ (a), and $MoO_3$ and 1L-$WSe_2$ when they are in contact (b). 1L-$WSe_2$ is degenerate p-doped due to the surface charge transfer at the $MoO_3$/1L-$WSe_2$ interface. (c-d) Low temperature photoluminescence (PL) spectra of as-prepared 1L-$WSe_2$ (c) and $MoO_3$/1L-$WSe_2$ heterostructure (d). The spectra were taken at $T$ = 80 K. X, X$^-$, and X$^+$ represent neutral excitons, negative trions, and positive trions,

respectively. The insets show the schematic of negative trions (electron plus exciton) and positive trions (hole plus exciton). The $MoO_3$/1L-$WSe_2$ heterostructure exhibits a strong PL signal from positive trions, suggesting a prevalent p-type doping due to the high electron affinity $MoO_3$.

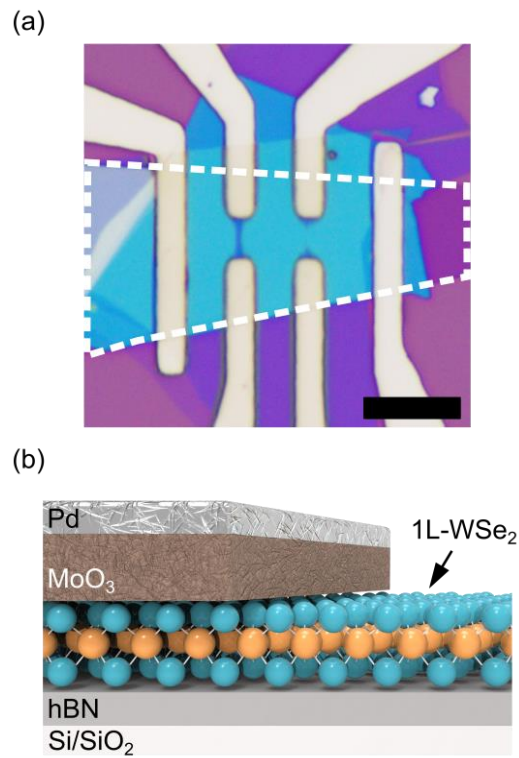

**Figure 2.** (a) Optical micrograph of a $MoO_3$ contacted 1L-$WSe_2$ FET with multiple electrodes. The 1L-$WSe_2$ nanoflake is indicated by the white dash line. The channel length and width are 19 and 14.4 μm, respectively. The distance between four-probe voltage $V_{4p}$ probes is 7 μm. Scale bar: 10 μm. (b) A cross-sectional schematic of $MoO_3$ contacted 1L-$WSe_2$ FETs.

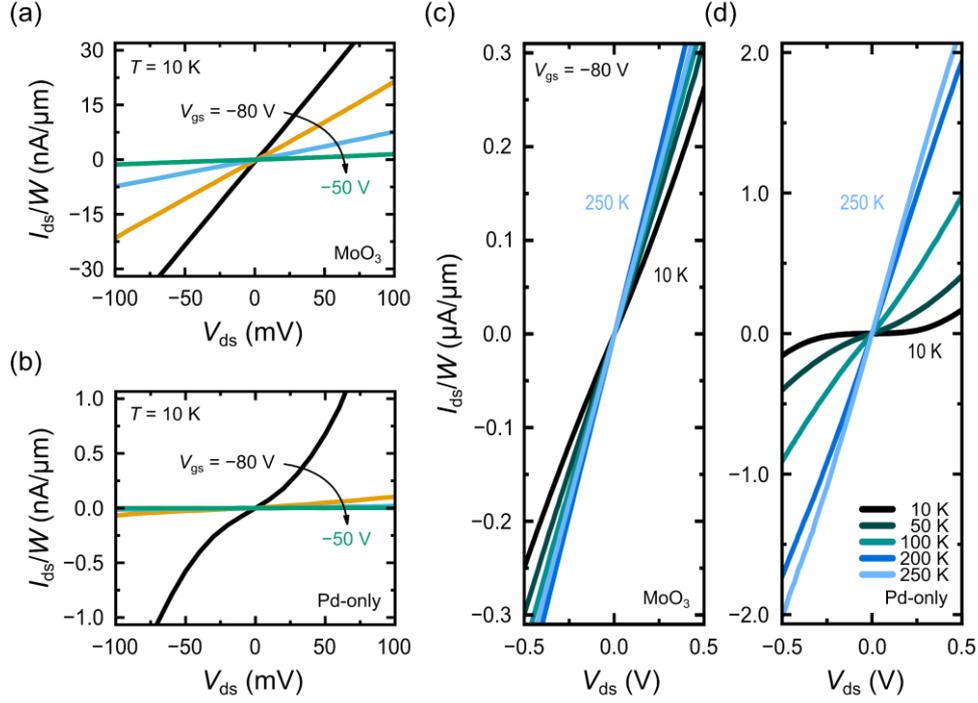

**Figure 3.** Transport characteristics of 1L-WSe$_2$ FETs with MoO$_3$ and Pd-only contacts. (a-b) Output curves at $T = 10$ K and gate voltage $V_{gs}$ ranging from −80 V to −50 V with an increment of 10 V for 1L-WSe$_2$ FETs with MoO$_3$ contacts (a) and Pd-only contacts (b). (c-d) Temperature-dependent output curves at gate voltage $V_{gs} = -80$ V for 1L-WSe$_2$ FETs with MoO$_3$ contacts (c) and Pd-only contacts (d).

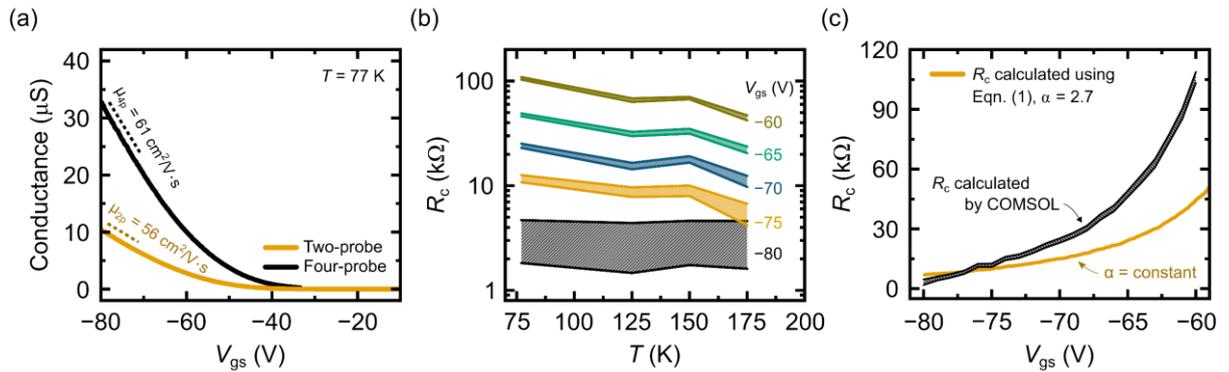

**Figure 4.** Transport characteristics of a MoO$_3$ contacted 1L-WSe$_2$ FET. (a) Drain current $I_{ds}$ as a function of gate voltage $V_{gs}$ at $T = 77$ K obtained by two- and four-terminal measurement. Drain voltage $V_{ds} = 1$ V (b) Calculated contact resistance $R_c$ as a function of temperature at gate voltage $V_{gs}$ ranging from −60 V to −80 V. (c) Contact resistance $R_c$ as a function of gate voltage

$V_{gs}$ at $T$ = 77 K calculated by COMSOL (grey shaded area) and assuming a constant α, α = 2.7 (orange solid line).

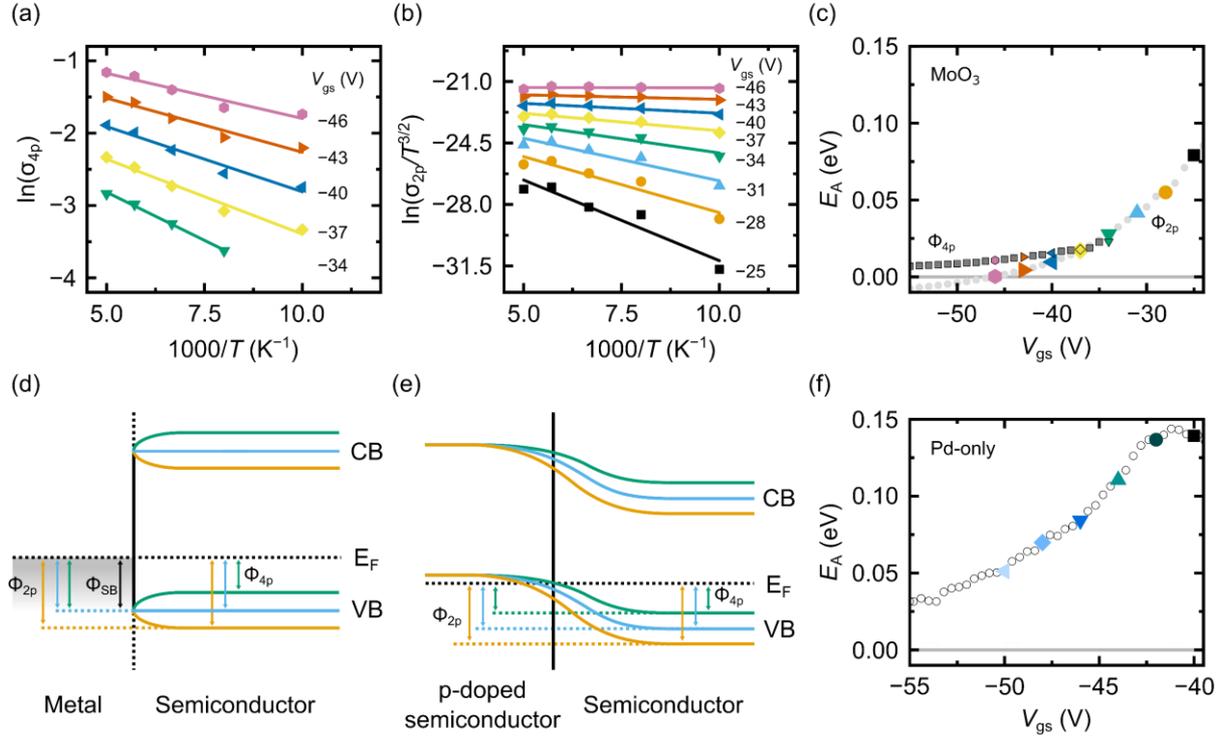

**Figure 5.** (a-b) Arrhenius plots of $\sigma_{4p}$ (a), $\sigma_{2p}/T^{3/2}$ (b) for a 1L-WSe$_2$ FET with MoO$_3$ contacts. $\sigma_{2p}$ is two-probe conductivity, $\sigma_{4p}$ is four-probe conductivity, and $T$ is the temperature. (c) Inferred activation energies $E_A$ of bulk channel and contact, $\Phi_{4p}$ and $\Phi_{2p}$ respectively, as functions of gate voltage $V_{gs}$ for the MoO$_3$ contacted 1L-WSe$_2$ FET extracted from (a, b). (d, e) band diagram for metal/semiconductor interface (d) and the interface between p-doped semiconductor and semiconductor (e), showing schematically the variation of $\Phi_{4p}$ and $\Phi_{2p}$ with gate voltage (different coloured lines; orange more positive $V_{gs}$, blue intermediate $V_{gs}$, green more negative $V_{gs}$). (f) Inferred activation energies $E_A$ as functions of gate voltage $V_{gs}$ for a 1L-WSe$_2$ with Pd-only contacts.